\documentstyle[multicol,prl,epsf,aps]{revtex}

\begin{document}
\draft
\title{Parametric Generation of Second Sound by First Sound in
Superfluid Helium} 
\author{Dmitry Rinberg, Vladimir Cherepanov, and Victor Steinberg} 
\address{Department of Physics of Complex Systems, The Weizmann  
Institute of Science, 76100 Rehovot, Israel}
\date{\today }
\maketitle

\begin{abstract}
We report the first experimental observation of parametric generation
of second sound (SS) by first sound (FS) in superfluid helium in a
narrow temperature range in the vicinity of $T_\lambda $. The
temperature dependence of the threshold FS amplitude is found to be in
a good quantitative agreement with the theory suggested long time ago
\cite{Pokrovskii-76} and corrected for a finite geometry. Strong
amplitude fluctuations and two types of the SS spectra are observed
above the bifurcation. The latter effect is quantitatively explained
by the discreteness of the wave vector space and the strong
temperature dependence of the SS dissipation length.
\end{abstract}


\begin{multicols}{2}
\narrowtext

Parametric generation of waves is observed in a wide class of
nonlinear media. Spin-waves in ferrites and antiferromagnets and
Langmuir waves in plasma parametrically driven by a microwave field,
ferrofluid surface waves subjected to an ac tangential magnetic field,
and surface waves in liquid dielectrics parametrically excited by an
ac electric field are just a few examples (see, e.g.,
\cite{S,L'vov_b-94,Cross-93}). Parametric excitation of standing
surface waves by vertical vibration is another canonical example of
such a system which recently has become particularly popular due to
easy visualization of its pattern dynamics. Depending on the number of
nonlinear modes, either nonlinear patterns or a wave turbulent state
are observed. In the case of the Faraday instability, a rather modest
experimentally reachable aspect ratio between the system's horizontal
size and the wavelength, and a relatively large dissipation, strongly
limit advantages of this system to study the wave turbulence which is
expected to exhibit some rather universal properties
\cite{L'vov_b-94}. From this point of view, the spin-wave instability
seems to present a better example of a system with many excited modes,
thanks to its extremely large aspect ratio and minute
dissipation. However, in spite of a well developed and detailed theory
for this system, the few experimental detection techniques accessible
in this case limit quantitative investigation of the wave turbulent
state\cite{L'vov_b-94}. 

In this Letter, we report the first experimental observation of
parametric generation of second sound (SS) waves by first sound (FS)
waves in superfluid helium in a rather narrow temperature range close
to the superfluid transition temperature $T_\lambda $. Due to
relatively weak dissipation and a large attainable aspect ratio on the
one hand, and experimental advantages of using high sensitivity
amplitude measurements on the other hand, this system may become an
appropriate one to quantitatively study the wave turbulence.
However, we are aware of only one experiment on nonlinear (not
parametric) interaction of first and second sounds \cite{Garrett-77}.

In superfluid helium, there are two kinds of three-wave processes that
are responsible for SS generation by FS waves: \v{C}erenkov emission
and parametric decay. These processes were considered theoretically
about 20 years ago \cite{Pokrovskii-76,Pushkina-74}. In the first one,
a FS phonon decays into a pair of FS and SS phonons. In the second
process, a FS phonon decays into a pair of SS phonons. Both processes
result in decay instabilities which are characterized by thresholds in
the FS amplitude. In this paper, we concentrate on the parametric
instability. 

In the parametric excitation process, a FS wave with wave vector 
${\bf K}$ and frequency $\Omega $ decays into two SS waves with wave
vectors ${\bf k_1}$ and ${\bf k_2}$ and frequencies $\omega _2$ and
$\omega _2$ that obey the conservation laws (resonance conditions) 
\begin{equation}
\Omega = \omega _1 + \omega _2,   \;\;
{\bf K} = {\bf k_1} + {\bf k_2}.  
\label{forCons}
\end{equation}
$\Omega = c_1 K$, $\omega _{1,2} = c_2\,k_{1,2}$, and $c_1$ and $c_2$
are the first and second sound velocities. A FS wave of amplitude $b$
generates SS waves at the rate $|bV|$, where $V$ is the interaction
matrix element \cite{Pokrovskii-76}. The SS waves dissipate at the
rate $\gamma $. The parametric instability occurs when the SS
generation exceeds the dissipation. The threshold FS amplitude is
$b_{th} = \gamma /\left| V_{\max} \right| $ \cite{Pokrovskii-76}. 

It was found in Ref. \cite{Pokrovskii-76} that in an infinite system,
the temperature range where the parametric excitation occurs before
the \v{C}erenkov emission is between 0.9 K and 1.2 K. However, our
estimates, based on the asymptotic behavior of the thermodynamic
properties of superfluid helium near $T_\lambda $ \cite{Ahlers-76},
show that the decay instability threshold is almost constant in a wide
range of reduced temperature $ 10^{-6} < \varepsilon < 10^{-2}$, where
$\varepsilon = (T_\lambda - T)/T_\lambda $. The threshold of the
\v{C}erenkov emission in the same range is higher than that of the 
parametric instability and diverges at $T_\lambda $ as
${\varepsilon}^{-0.8}$. Thus this region is convenient to conduct
experiments. In this temperature range, the matrix element $V$ reaches
its maximum value for SS waves propagating perpendicularly to the FS 
wave. Moreover, since the attenuation length of the SS, $l$, varies
drastically in this temperature range \cite{Mehrotra-84}, the finite
length of the cell in the direction of the SS propagation should be
taken into account. 

\begin{figure}[tbp]
\epsfxsize = \hsize
\epsffile{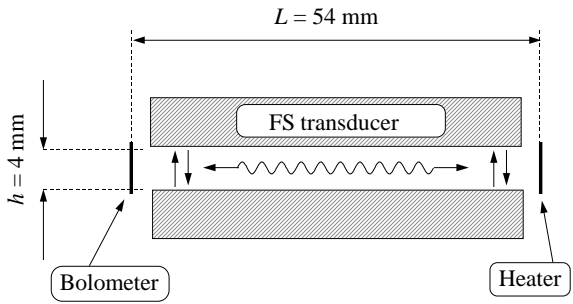}
\caption{The experimental cell and the geometry of FS and SS wave
propagation.}
\label{figCell}
\end{figure}

The experiment was done on a short cylindrical cell of 50~mm in
diameter and length $h=4$ mm (Fig \ref{figCell}). The sides of the
cavity were formed by a pair of capacitor transducers. Such
transducers and their calibration method are described in
Ref.\cite{Barmatz-68}. In the short direction, the cell was used as a
FS resonator with the first harmonic frequency about 28~kHz . The
frequency of parametrically excited SS waves was about 14~kHz. In the
experimentally investigated temperature range, $10^{-4} < \varepsilon
< 2\times 10^{-3}$ the SS wavelength $\lambda $ varied from 0.09 to
0.3~mm. The resonator's lateral side was open, and several layers of
paper were put around it to absorb transverse acoustic modes. The
resonator quality factor was $Q\approx 150$: that allowed one to
neglect influence of all other acoustic modes.

Two SS bolometers and two heaters were evaporated on 22 mm diameter
glass substrates. One bolometer was placed diametrically opposite to
each heater and the two pairs were mounted at $90^{\circ }$ to each
other. The bolometers were 40~$\mu $m wide superconducting Au-Pb line
in the form of a round serpentine pattern of diameter $d_b=2$~mm
\cite{Hanan-92}. Each heater--bolometer pair, with the $L=54$~mm base,
formed a resonance open cavity for the SS, with resonances from 12 to
40~Hz apart. Possible incidental effects of these resonances on
experimental observations are discussed below. The bolometers were
sensitive to waves incident within the angle $\sim \lambda /d_b$ to
the normal direction, i.e., from 0.05 to 0.15 for $\varepsilon $ from
$10^{-4}$ to $2\times 10^{-3}$. The cell was placed into a container
filled with about 300 ml of purified He$^4$ and vacuum sealed. A
three-stage temperature regulated cryostat with the helium container
as the third stage was used \cite{Hanan-92,Steinberg-83}. Its
temperature stability was about $10^{-6}$ K, and the operating
temperature range was between 1.9 K and $T_\lambda $.

The experiment was performed at a constant temperature by generating
FS at pump frequency $F_p$, which was chosen to be the first harmonic
frequency of the cavity. For a given FS amplitude, the SS amplitude
was measured by the bolometers. In a typical experimental run, a
signal from a bolometer was sent to a preamplifier and then to a
lock-in-amplifier at reference frequency $F_p/2$. The two-phase output
signal was sampled at 16~Hz and digitized over a 128~s period. The
complex Fourier transform of the both digitized components of the
signal provided the SS spectrum in a narrow window, $\pm 8$~Hz, around
$F_p/2$, with resolution of $1/128$~Hz. 

\begin{figure}[tbp]
\epsfxsize = \hsize
\epsffile{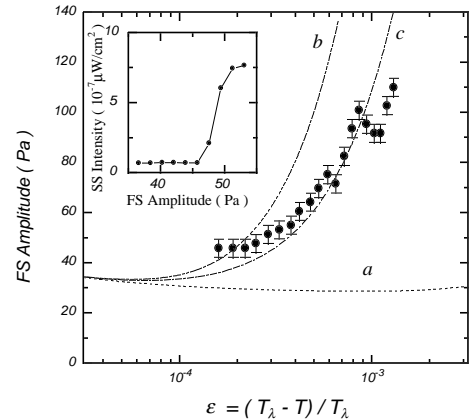}
\caption{The temperature dependence of the parametric excitation
  threshold (FS amplitude): curve ($a$) -- theory for infinite cell
  ($r = 1$) \protect\cite{Pokrovskii-76}, curves ($b$) and ($c$) --
  theory for finite cell ($r = 0$) and ($r = 0.6$), solid circles with
  error bars -- experimental data. Inset -- a typical plot of SS
  intensity versus the FS amplitude in the vicinity of the
  threshold. } 
\label{figThresh}
\end{figure}

A typical plot of the time averaged SS intensity as a function of the 
driving FS amplitude in the vicinity of $F_p/2$ is shown on the inset
of Fig.~\ref{figThresh}. There exists a well pronounced threshold for
the driving amplitude at which the SS intensity first exceeds twice
the background noise level. We point out that the SS amplitude as a
function of time strongly fluctuates. Its intermittent behavior is
particularly striking close to the onset where the time intervals
between isolated spikes increases up to 1,000 seconds which was the
maximum sampling interval. This time scale is much longer than all
characteristic scales in the problem. Those fluctuations limit the
resolution in the threshold determination and hinder quantitative
studies of the amplitude behavior above the onset. Nevertheless, they
do not change the main features of the phenomena discussed in the
paper. The threshold FS amplitude as a function of temperature is
shown in Fig.~{\ref{figThresh}} together with theoretical predictions. 

As mentioned above, heater--bolometer pairs, $L = 54$~mm apart, formed  
boundaries for SS waves. In our experiments, the attenuation length
$l$ changed from 10 to 600~mm. Therefore, reflections from the
boundaries play an important role in the parametric instability. A
straightforward generalization of the envelope method
\cite{L'vov-77,CS-rev} for a system of the size $L$, where waves
arrive and reflect normally to the boundaries, with an arbitrary
reflection coefficient $r$, yields the following expression for the
threshold  
\begin{equation}
b_{th} = ( \gamma /V) \left[ 1 + 
{\xi}^2 \left( l/L \right) ^2 \right] ^{1/2},
\label{forThresh}
\end{equation}
where $\xi $ is the minimal positive root of the equation 
\begin{equation}
\tan \xi = - \frac{\left( 1 - r^2 \right) \xi (l/L)}{ 1 + r^2 
- 2 r \left[ 1 + {\xi}^2 \left( l/L \right) ^2 \right] ^{1/2}}.  
\label{forXi}
\end{equation}
For $L < l,$ the value of $b_{th}$ (along with its temperature
dependence) is mainly determined by the second term in the parenthesis
in Eq.~(\ref{forThresh}), i.e., $b_{th}\sim c_2/(LV)\propto
\varepsilon ^{1.33}$ (see curve ($b$) in Fig.~\ref{figThresh}). Note
that if the reflection coefficient tends to unity, the threshold
coincides with that in an infinite system. 

To compare results of our threshold measurements with theoretical 
predictions we plotted theoretical curves for the cases of perfect 
reflection ($r=1$) - curve ({\it a}) in Fig.~{\ref{figThresh}}, and of
the complete absorption ($r=0$) - curve ({\it b}) of SS on the
boundaries. The SS dissipation rate, $\gamma $, was found from
experimental measurements \cite{Mehrotra-84}. The curves converge in
the region $\varepsilon < 2 \times 10^{-4}$, where the size effect is
negligible ($l < L$). We also plot a curve ({\it c}) for the
reflection coefficient $ r = 0.6$. In the temperature range $4 \times
10^{-4}<\varepsilon < 2 \times 10^{-3}$, this curve fits the
experimental data rather well. Near the edges of the cell, SS waves
are not amplified but they do attenuate. With approaching $T_\lambda
$, the attenuation length, $l$, becomes comparable to the
characteristic length of the edge effects, $h$. This results in a
decrease of reflection from the cell boundaries. Therefore the
effective reflection coefficient decreases and the experimental points
lie closer to the theoretical curve ({\it b}) in the temperature range
$\varepsilon < 4 \times 10^{-4}$. Because of a steep decrease of the
amplitude of the measured signal we did not succeed to extend our
measurements closer to $T_\lambda $ than $10^{-4}$. 

\begin{figure}[tbp]
\epsfxsize = \hsize
\epsffile{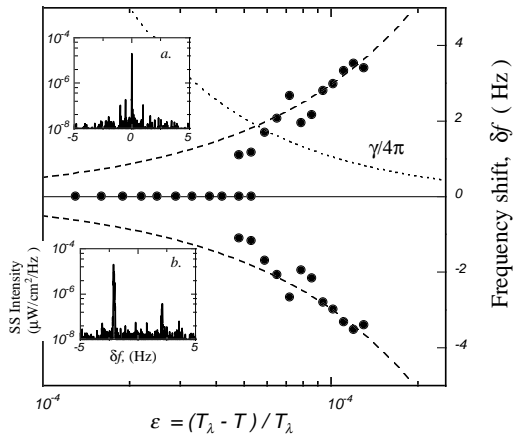}

\caption{The SS frequency shift as a function of $\varepsilon $. The
  insets: two types of the SS power spectra in two temperature
  ranges: ($a$) $\varepsilon = 2.19 \cdot 10^{-4}$, ($b$) $\varepsilon
  = 6.52 \cdot 10^{-4}$.}
\label{figSpect}
\end{figure}

Another result, which was unexpected for us, was the observation of 
different types of SS power spectra in different temperature
regions. Two typical spectra of the SS signals close to the onset at
the reduced temperatures $\varepsilon = 2.19 \times 10^{-4}$ and
$\varepsilon = 6.52 \times 10^{-4}$ are shown in the insets of
Fig.~\ref{figSpect}. The first spectrum has a single sharp peak at
exactly $F_p/2$, while the second one, observed further from
$T_\lambda $, exhibits two equidistant peaks around $F_p/2$. The left
peak at the lower frequency $F_p/2 - \delta \!f$ is always larger than
the right one at the frequency $F_p/2 + \delta \!f$.

\begin{figure}[tbp]
\epsfxsize = \hsize
\epsffile{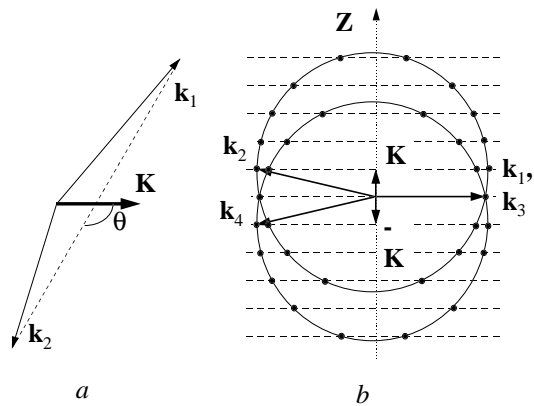}
\caption{($a$) $k$-vector diagram for infinite space, ($b$) the
  discrete wave vector space and the geometry of FS and SS interaction
  in a finite cell. }
\label{figVect}
\end{figure}

In order to explain such a spectrum splitting one needs to consider
the conservation laws (\ref{forCons}). In the temperature range where
our experiment was performed ($10^{-4} < \varepsilon < 2 \times
10^{-3}$), the sound velocity ratio $\eta =c_2/c_1$ varies between
0.06 and 0.019 \cite{Ahlers-76}. From the conservation laws
(\ref{forCons}) one gets $K\ll k_1,k_2$, i.e., the SS wave vectors are
located on an ellipsoid of revolution with an eccentricity of $\eta $,
and ${\bf k_1}$ and ${\bf k_2}$ are almost opposite
(Fig.~\ref{figVect}a). A more detailed analysis of these equations
reveals an angular dependence of the SS frequency  
\begin{eqnarray}
\omega _{1,2} = \Omega \left( 1 \pm \nu \right) /2, \;\; 
\nu = \eta \cos \theta ,  
\label{forShift} \\
\cos \theta  = \frac{\left( {\bf k}_1 - {\bf k}_2 \right) 
\cdot {\bf K}}{ \left| {\bf k}_1 - {\bf k}_2 \right| 
\left| {\bf K} \right| }.  
\label{forAngle}
\end{eqnarray}
A weak angular dependence of the threshold amplitude $b_{th}$ due to
an angular dependence of $V$ \cite{Pokrovskii-76} yields a weak
minimum for the symmetric configuration with $\theta = \pi /2$.

The existence of two types of spectra and the transition between them
can be explained by the finite cell geometry in the direction of FS
propagation and by discreteness of the wave vector in this direction
(see Fig.~\ref{figVect}b). Indeed, if the attenuation rate $\gamma $
is larger than the frequency difference between resonance peaks,
$\Delta \omega = 2 \left( 2\pi \delta \!f \right) = \Omega \nu $, that
occurs closer to $T_\lambda $, the discreteness of the resonance
states is smeared out. Then just a single peak appears in the
spectrum. This symmetric case corresponds to the minimum of the
threshold as a function of the angle $\theta $ for the infinite cell
geometry\cite{Pokrovskii-76}. The opposite inequality $\Delta \omega
> \gamma $ corresponds to discrete resonance states and leads to two
peaks in the power spectra. 
If $\Delta \omega $ is sufficiently large compared to $\gamma $, one
can consider two parametric excitation processes caused by two
components of the FS standing wave, with the wave vectors $\pm {\bf
  K}$. The momentum conservation conditions for these processes are  
\begin{equation}
{\bf K} = {\bf k_1} + {\bf k_2}, \;\;
{\bf -K} = {\bf k_3} + {\bf k_4}.
\label{forMoment}
\end{equation}
The processes do not interfere unless the same SS phonon participates
in both processes. If that happens, e.g., ${\bf k_1} \equiv {\bf
  k_3}$, the two components of the standing wave contribute to the
excitation process. Therefore, this process has a lower threshold. It
follows from Eqs.~(\ref{forShift}--\ref{forMoment}) that in this case
the common wave vector ${\bf k_1}$ is exactly perpendicular to the FS
wave vector ${\bf K}$, the waves with ${\bf k_1}$, having the lower
frequency, $\cos \theta = \eta $, and $\nu = {\eta }^2$. The splitting
transition in the SS spectrum occurs when ${2} \left( {2\pi } \delta
\!f \right) /{\gamma } \approx 1$.

The experimental data on the frequency shift and the attenuation rate
$\gamma /4\pi $ (dotted line) are plotted in
Fig.~\ref{figSpect}. Closer to $T_\lambda $, the spectrum has just one
peak at the frequency $F_p/2$ that corresponds to the symmetric decay
of FS. Further away from $T_\lambda $ ( $\varepsilon > 6 \times
10^{-4}$ ), the spectrum changes to two peaks equally separated from
$F_p/2$. The crossover from one type of the spectrum to another occurs
in the reduced temperature interval $4 \times 10^{-4} < \varepsilon <
6 \times 10^{-4}$ where both the central peak and a pair of separated
peaks coexist in the spectrum. The temperature dependence of the
frequency shift of the peaks is described rather well by $\delta \!f =
\pm (F_p/2){\eta }^2$ (dashed line). This corresponds to the decay
process in which one of the SS waves propagates normally to the FS
direction, and both components of the FS standing wave contribute to
the parametric excitation. 

The large difference between the peak amplitudes can be explained as a
result of constructive interference of SS waves propagating exactly in
the cell plane. The smaller peak at the higher frequency corresponds
to two SS waves which propagate at a small angle to the FS waves and
interfere destructively.

Two heater--bolometer pairs allowed us to examine the SS correlations
on the bolometers. We found no significant correlations between the
signals on the bolometers near the threshold. That indicates no
tendency to pattern formation. The reason for such an essential
difference from surface wave experiments is probably as follows:
normal modes were not generated in our partially open cell that
changed completely the pattern formation conditions compared to those
in the Faraday crispation experiments \cite{Douady-88}. 

In conclusion, a number of new features distinguish the SS parametric 
excitation in superfluid He from well studied parametric instabilities
of spin-waves in magnets and of surface waves on liquid: (i) the
instability was observed in a partially open system, (ii) the
finiteness of the system lifts the degeneracy in pairs of
parametrically excited SS waves that leads to the spectrum splitting
in the temperature range $\varepsilon > 5 \cdot 10^{-4}$, and (iii)
the high sensitivity of the setup allowed us to observe strong
fluctuations of the SS amplitude very close to the threshold. 

We gratefully acknowledge useful and helpful conversations with
V.~L'vov and experimental advice by M.~Rappaport and M.~Reznikov. 

This work was partially supported by a grant from the Minerva
Foundation and by the Minerva Center for Nonlinear Physics of Complex
Systems.

\end{multicols}{2}


\begin{references}

\bibitem{Pokrovskii-76}  I.~M.~Khalatnikov and V.~L.~Pokrovskii, 
{Sov. Phys. JETP} {\bf 44}, 1036 (1976).

\bibitem{S}  H.~Suhl, {\it J. Phys. Chem. Solids} {\bf 1}, (1957) 209. 

\bibitem{L'vov_b-94}  V.~L'vov, {\em Wave Turbulence Under Parametric 
Excitation. (Applications to Magnetics)} (Springer-Verlag, 1994).

\bibitem{Cross-93}  M.~C.~Cross and P.~C.~Hohenberg, {Rev. Mod. Phys.}
 {\bf 65}, 851 (1993).

\bibitem{Garrett-77}{S.~L.~Garrett, Technical  Report No. 39,
    Department of Physics, UCLA, (1977).} 

\bibitem{Pushkina-74}  N.~I.~Pushkina and R.~V.~Khokhlov, 
{Sov. Phys. JETP Lett.} {\bf 19}, 348 (1974).

\bibitem{Ahlers-76}  G.~Ahlers, in {\em The Physics of Liquid and
Solid Helium: Part I}, edited by K.~H.~Bennemann and J.~B.~Ketterson
(John Wiley and Sons, Inc., N.Y. , New York, 1976).

\bibitem{Mehrotra-84}  R.~Mehrotra and G.~Ahlers, {
Physical Review B} {\bf 30}, 5116 (1984).

\bibitem{Barmatz-68}  I.~Rudnick and M.~Barmatz, 
{Phys.Rev.} {\bf 170}, 224 (1968).

\bibitem{Hanan-92}  H.~Davidowitz, PhD thesis (The Feinberg Graduate
  School of the Weizmann institute of Science, 1992).

\bibitem{Steinberg-83}  V.~Steinberg and G.~Ahlers, 
{J. Low Temp. Phys.}  {\bf 53}, 255 (1983).

\bibitem{L'vov-77}  V.~S. L'vov and A.~M. Rubenchik, 
{Sov. Phys. JETP} {\bf 45}, 67 (1977).

\bibitem{CS-rev}  V.~B.~Cherepanov and A.~N.~Slavin in 
{\em High frequency processes in magnetic materials}, edited by
G.~Srinivasan and A.~N.~Slavin (World Scientific, 1995).

\bibitem{Douady-88}  {S.~Douady and S.~Fauve, Europhys. Lett. 
{\bf 6}, 221 (1988), 
A.~B.~Ezerskii, M.~I.~Rabinovich, V.~P.~Reutov, and
I.~M.~Starobinets, Sov. Phys. JETP {\bf 64}, 1228 (1986).}
\end{references}
\end{document}